\documentclass[aps,preprint]{revtex4}
\usepackage{amsmath}
\usepackage{blindtext}
\usepackage{extarrows}
\usepackage{braket}
\usepackage{pdfpages}
\usepackage{graphicx}
\usepackage{caption}
\usepackage{float}
\graphicspath{ {images/} }
\usepackage{amsmath,graphicx,color,xcolor,epsfig}
\usepackage{epstopdf}
\usepackage{hyperref}

\providecommand{\U}[1]{\protect\rule{.1in}{.1in}}

\begin{document}
	
	\baselineskip 18pt
	\parindent 12pt
	\parskip 10pt
	\preprint{}
	\title{\textbf{Dynamics of Work Extraction in Multipartite Atomic Systems: Role of Correlations and Relative Entropy}}
	\author{M. Ibrahim}
	\email{ibrahim@phys.qau.edu.pk}
	\author{S. J. Anwar}
	\author{K. Khan}
	\affiliation{Department of Physics, Quaid-i-Azam University, Islamabad, Pakistan}
	
	\date{\today }
	
	\begin{abstract}
 We investigate the dynamics of quantum ergotropy in a multipartite system of two-level atoms interacting with a quantized field. The role of multipartite quantum mutual information and quantum relative entropy is analyzed to understand how correlations and the distinguishability between the passive state and the product of local states affect the extractable work. Numerical results for systems containing two to five atoms show that the maximum ergotropy does not increase linearly with the system size, while the intervals of zero ergotropy are gradually modified and suppressed for larger systems. We also study the effect of the average thermal photon number and different initial atomic states. Increasing the thermal photon number mainly reduces the relative-entropy contribution, whereas the initial state strongly influences both the magnitude and persistence of ergotropy. Among the considered states, the partially entangled state provides the largest and more persistent extractable work, while the mixed GHZ state gives the smallest ergotropy. These results highlight the importance of multipartite correlations, thermal effects, and state preparation in controlling quantum work extraction and may be useful for future studies of quantum batteries and quantum information resources.

	\end{abstract}
	
	\maketitle
	
	\section{\qquad Introduction}
	
	Over the past 20 years, thermodynamics and quantum information theory has become one of the most productive areas for fundamental physics. Understanding the thermodynamic behavior of systems functioning in the quantum world, where coherence, correlations, and entanglement are crucial, is the goal of this research, which is frequently referred to as quantum thermodynamics \cite{Goold2016, Lostaglio2019}. Ergotropy, which is the maximum amount of work that can be extracted from a quantum state through unitary cyclic operations. Ergotropy was first defined by Allahverdyan et al. \cite{Allahverdyan2004}, and it is now a fundamental concept for examining the thermodynamic value of quantum resources. Our knowledge of its boundaries and operational importance, including basic constraints imposed by quantum fluctuations and the structure of the underlying Hilbert space, has been improved by later research \cite{Perarnau2017}. The term "passive" refers to a state from which no work can be extracted. The passive state associated with a density matrix for a particular Hamiltonian is its least-energy representation in the eigenbasis \cite{Pusz1978, Lenard1978}. The ergotropy is then the energy difference between the original state and this passive counterpart.
	
	A particularly compelling line of inquiry has been the connection between ergotropy and quantum coherence, as well as more general quantum correlations \cite{Streltsov2017, Baumgratz2014}. It has been shown that the extractable work from a quantum system is intimately linked to its coherence in the energy eigenbasis, and that the amount of ergotropy can be decomposed into contributions arising from coherent superpositions and populations \cite{Francica2017}. This decomposition has led to the concept of \textit{daemonic} ergotropy, where quantum correlations with an ancilla can enhance the work that can be extracted locally \cite{Francica2020}. Moreover, a powerful information-theoretic framework for comprehending ergotropy is offered by the quantum relative entropy, which quantifies the distinguishability between two quantum states. Indeed, a formal relationship between thermodynamic work and information geometry may be established by expressing the ergotropy as the smallest relative entropy between the state and any passive state with the same spectrum \cite{Sone2021, Vedral2002}.
	
	Concurrently, one of the main challenges in multipartite quantum systems is the quantification of correlations. The categorization and measurement of entanglement in systems with three or more parties is significantly more difficult, although bipartite entanglement is comparatively well understood. The existence of inequivalent classes of entangled states under stochastic local operations and classical communication, such the classic Greenberger-Horne-Zeilinger (GHZ) and W states \cite{Greenberger1989, Dur2000}, is the cause of this complexity. Entanglement witnesses, entropy-based criteria, and geometric metrics are just a few of the methods that have been developed over time to identify and assess true multipartite entanglement \cite{Horodecki2009, Guhne2009}. Nevertheless, these techniques are frequently computationally intensive for particular state families. This has initiated efforts to find other tests that are operationally accessible, like thermodynamic quantities \cite{Amico2008, Eisert2010, Laflorencie2016}. Ergotropy has recently been shown to be an effective thermodynamic probe of multipartite entanglement. For instance, the difference between global and local ergotropies has been suggested as a faithful measure to identify distinct entanglement classes and capture the signature of true multipartite entanglement, providing an experimentally viable way to observe non-classical correlations through energy measurements \cite{SPP2022}.
	
	In this context, the Tavis-Cummings (TC) model offers a perfect framework for these kinds of studies \cite{Tavis1968, Dicke1954}. Superradiance, collapses and revivals, and the creation of complicated multipartite entangled states are just a few of the rich phenomena that arise from its dynamics, which are usually investigated under the rotating wave approximation \cite{Shore1993}. The TC model is an ideal testbed for examining the interaction between multipartite correlations and thermodynamic variables because of the collective nature of the atom-field coupling. Its importance for testing basic ideas in quantum thermodynamics and information is further enhanced by its experimental realizability across a variety of platforms, including trapped ions, circuit QED, and cold atomic ensembles \cite{Walborn2006, Lloyd2011}.
	
	Despite these significant developments, a number of crucial issues are still unresolved. The majority of research on correlations and ergotropy has focused on bipartite systems or particular groups of multipartite states. There is currently no systematic study of the \textit{dynamical} evolution of multipartite ergotropy and its relationship to other basic information-theoretic metrics such as the quantum relative entropy and the quantum mutual information \cite{Alicki2013, Barra2015}. Understanding how information spreads and is used during work extraction protocols depends on the mutual information's dynamical behavior under collective interactions, which gives a measure of the total correlations, including both classical and quantum parts \cite{Skrzypczyk2014, Huber2015}. Furthermore, an exact measure of how thermodynamically "active" a state is at any given moment is the relative entropy between the dynamical state and its immediate passive equivalent. It is yet unclear how various initial atomic states, including pure GHZ states, mixed (Werner-like) GHZ states, and other entangled states, affect the dynamical generation and decay of these thermodynamic resources \cite{Hovhannisyan2013}. Additionally, although the impact of temperature on ergotropy has been investigated in certain situations, its impact on the dynamics of these multipartite thermodynamic variables in a system with collective interactions has not been thoroughly described. The available extractable work and the underlying correlation structure can be drastically changed by temperature variations \cite{Santos2019, Levy2020}.
	
	Motivated by these unsolved issues, we present a thorough analysis of the dynamics of quantum relative entropy, quantum mutual information, and multipartite. In particular, we study the systems of two to five two-level atoms interacting with a single-mode cavity field. To clarify the influence of initial correlations and entanglement on work extraction and information dynamics, we examine the dynamical evolution of these parameters for a broad range of initial atomic states, such as pure GHZ states, mixed GHZ states, and other entangled states. Furthermore, by considering thermal states for the atomic ensemble, we expand our analysis to finite temperatures and investigate how thermal excitations impact the extractable work and the effectiveness of turning correlations into thermodynamic fuel. We seek to provide a better understanding of the thermodynamic and informational features of quantum systems by examining and uncovering the relation between multipartite ergotropy, quantum mutual information, and relative entropy. We describe in detail the extensive dynamical behavior revealed by our results, which include a non-trivial reliance on system size and temperature and unique signatures of entanglement classes in the evolution of ergotropy \cite{Sharma2025, Nielsen2010}.
	
	The paper is organized as follows. In Sec. II, we introduce the Tavis-Cummings Hamiltonian and describe the dynamics of the system. Section III defines the central quantities of our study: multipartite ergotropy, multipartite quantum mutual information, and the quantum relative entropy with respect to the passive state. Our main results on the dynamics of these quantities for different system sizes and initial states are presented and discussed in Sec. IV, including the effects of temperature. Finally, Sec. V concludes with a summary and outlook.
	\section{Hamiltonian Model}
In the present study, we employ the Tavis--Cummings model
\cite{Tavis1968} to describe the interaction between a multipartite atomic
subsystem and a single quantized cavity mode. The conventional
Tavis--Cummings model extends the Jaynes--Cummings interaction to an ensemble
of identical two-level atoms that are collectively coupled to the same
radiation field. The dynamical behavior is investigated for atomic ensembles
consisting of two, three, and four two-level atoms.

The Hamiltonian governing the interaction of $N$ identical two-level atoms
with the cavity field is given by
\begin{equation}
	\begin{split}
		\hat{H}_{T}
		={}&
		\frac{\hbar\omega_{0}}{2}
		\sum_{i=1}^{N}\hat{\sigma}_{i}^{z}
		+
		\hbar\omega\hat{a}^{\dagger}\hat{a}
		\\
		&+
		\hbar g
		\sum_{i=1}^{N}
		\left(
		\hat{a}\hat{\sigma}_{i}^{+}
		+
		\hat{a}^{\dagger}\hat{\sigma}_{i}^{-}
		\right).
	\end{split}
	\label{total_Hamiltonian}
\end{equation}
Here, $\omega_{0}$ denotes the atomic transition frequency, while $\omega$
represents the frequency of the cavity mode. The operators
$\hat{\sigma}_{i}^{z}$, $\hat{\sigma}_{i}^{+}$, and $\hat{\sigma}_{i}^{-}$
denote the population-inversion, raising, and lowering operators of the
$i$th atom, respectively. The parameter $g$ represents the atom--field
coupling strength.

The initial state of the combined atom--field system is assumed to be
separable and is defined as the tensor product of a partially mixed
atomic state and a thermal cavity-field state:
\begin{equation}
	\hat{\rho}(0)
	=
	\hat{\rho}_{A}(0)
	\otimes
	\hat{\rho}_{F}^{\mathrm{th}}.
	\label{initial_total_state}
\end{equation}
The initial atomic density operator is defined as
\begin{equation}
	\hat{\rho}_{A}(0)
	=
	(1-p)|\psi_{A}\rangle\langle\psi_{A}|
	+
	p
	|g_{1}g_{2}\cdots g_{N}\rangle
	\langle g_{1}g_{2}\cdots g_{N}|,
	\label{initial_atomic_state}
\end{equation}
where $p$, with $0\leq p\leq 1$, determines the degree of mixedness of the
atomic subsystem. At $p=0$, the atoms are initially prepared in a pure
state, while increasing $p$ increases the contribution of the collective
ground state.

The pure multipartite atomic state is defined as
\begin{equation}
	|\psi_{A}\rangle
	=
	\cos(\theta)
	|g_{1}g_{2}\cdots g_{N}\rangle
	+
	\sin(\theta)
	|e_{1}e_{2}\cdots e_{N}\rangle,
	\label{pure_atomic_state}
\end{equation}
where $0\leq\theta\leq\pi$. The parameter $\theta$ determines the relative
contributions of the collective ground and excited components.

The thermal cavity mode is represented by a statistical mixture of
photon-number states, with its density operator given by
\begin{equation}
	\hat{\rho}_{F}^{\mathrm{th}}
	=
	\sum_{n=0}^{\infty}
	P_{n}
	|n\rangle\langle n|,
	\label{thermal_field_state}
\end{equation}
where the thermal photon-number distribution is
\begin{equation}
	P_{n}
	=
	\frac{\bar{n}^{\,n}}
	{(1+\bar{n})^{n+1}}.
	\label{thermal_distribution}
\end{equation}
Here, $\bar{n}$ represents the average number of thermal photons in the
cavity and is related to the field temperature $T$ through the
Bose--Einstein distribution
\begin{equation}
	\bar{n}
	=
	\frac{1}
	{\exp\left(\hbar\omega/k_{\mathrm{B}}T\right)-1},
	\label{mean_thermal_photons}
\end{equation}
where $k_{\mathrm{B}}$ denotes the Boltzmann constant. For $\bar{n}=0$, the
thermal field becomes the vacuum state, whereas larger values of $\bar{n}$
represent higher thermal excitation of the cavity mode.\\

The permitted basis states $\left\{|\psi_i\rangle\right\}$ are expressed as
\begin{equation}
	\left\{|\psi_i\rangle\right\}
	=
	|g_1g_2\ldots g_N,n+N\rangle,\,
	|e_1g_2\ldots g_N,n+N-1\rangle,\,
	\ldots,\,
	|e_1e_2\ldots e_N,n\rangle.
\end{equation}
Here, $|e_i\rangle$ denotes the excited state of the $i$th atom. The state
of the complete atom--field system at time $t$ is written as
\begin{equation}
	\hat{\rho}_{AF}(t)
	=
	\sum_{i,j}^{N}
	|\psi_i\rangle
	\langle\psi_i|
	\hat{\rho}(t)
	|\psi_j\rangle
	\langle\psi_j|.
\end{equation}
Furthermore, the density operator $\hat{\rho}(t)$ can be expressed in terms
of the eigenvalues and eigenvectors of the initial density matrix as
\begin{equation}
	\hat{\rho}_{AF}(t)
	=
	\sum_{i,j}^{N}
	\exp\left[-i(E_i-E_j)t\right]
	\langle\psi_i|\hat{\rho}(0)|\psi_j\rangle
	|\psi_i\rangle
	\langle\psi_j|.
\end{equation}
In this expression, $E_i$, $E_j$ and
$|\psi_i\rangle$, $|\psi_j\rangle$ denote the eigenvalues and
eigenvectors of the initial density matrix $\hat{\rho}(0)$, respectively.
The atomic subsystem is subsequently obtained by tracing over the field
degrees of freedom:
\begin{equation}
	\hat{\rho}_{T}(t)
	=
	\operatorname{Tr}_{F}
	\left[
	\hat{\rho}_{AF}(t)
	\right].
\end{equation}
\section{Dynamical Multipartite Quantum Ergotropy, Mutual Information, and Relative Entropy}
	For a multipartite system composed of $N$ two-level atoms, denoted by
	$A,B,C,\ldots$, the dynamical quantum ergotropy can be related to the total
	correlations through the quantum mutual information and the quantum
	relative entropy as \cite{linkergo}
	\begin{equation}
		\beta \epsilon(\rho(t))
		=
		I(A:B:C:\cdots)
		-
		D\left(
		P_{\rho}(t)\middle||
		\rho_A(t)\otimes\rho_B(t)\otimes\rho_C(t)\otimes\cdots
		\right),
		\label{eq:ergotropy_relation}
	\end{equation}
	where $\beta$ is the inverse temperature and $\epsilon(\rho)$ denotes
	the ergotropy per system. The quantum mutual information for the
	multipartite state is defined by
	\begin{equation}
		I(A:B:C:\cdots)
		=
		\sum_{X=A,B,C,\ldots}S(\rho_X(t))-S(\rho(t)),
		\label{eq:mutual_multipartite}
	\end{equation}
	where $\rho_X$ is the reduced density matrix of subsystem $X$ and
	\begin{equation}
		S(\rho)=-\operatorname{Tr}(\rho\ln\rho)
	\end{equation}
	is the von Neumann entropy. In Eq.~\eqref{eq:ergotropy_relation}, the relative entropy is evaluated
	between the passive state $P_{\rho}$ and the product of the reduced
	states by $D(\rho||\sigma)
	=
	\operatorname{Tr}(\rho\ln\rho)
	-
	\operatorname{Tr}(\rho\ln\sigma)$. Thus,
	\begin{equation}
		D\left(
		P_{\rho}\middle|\middle|
		\bigotimes_X\rho_X
		\right)
		=
		\operatorname{Tr}(P_{\rho}\ln P_{\rho})
		-
		\operatorname{Tr}
		\left[
		P_{\rho}
		\ln\left(\bigotimes_X\rho_X\right)
		\right].
		\label{eq:relative_multipartite}
	\end{equation}
	The passive state has the same eigenvalues as the original state and
	can be written as
	\begin{equation}
		P_{\rho}
		=
		\sum_i\lambda_i|e_i\rangle\langle e_i|,
		\label{eq:passive}
	\end{equation}
	where $\lambda_i$ are the eigenvalues of $\rho$, arranged in decreasing
	order with increasing energy, and $|e_i\rangle$ are the corresponding
	energy eigenstates. To simplify the second term in
	Eq.~\eqref{eq:relative_multipartite}, we write the spectral
	decomposition of the product of the reduced states as
	\begin{equation}
		\bigotimes_X\rho_X
		=
		\sum_k\beta_k|\phi_k\rangle\langle\phi_k|,
		\label{eq:product_spectral}
	\end{equation}
	where $\beta_k$ and $|\phi_k\rangle$ are its eigenvalues and
	eigenvectors, respectively. Consequently, the relative entropy becomes
	\begin{equation}
		\begin{aligned}
			D\left(
			P_{\rho}\middle|\middle|
			\bigotimes_X\rho_X
			\right)
			&=
			\sum_i\lambda_i\ln\lambda_i
			-
			\sum_{i,k}
			\lambda_i
			|\langle e_i|\phi_k\rangle|^2
			\ln\beta_k .
		\end{aligned}
		\label{eq:relative_final}
	\end{equation}
	Substituting Eqs.~\eqref{eq:mutual_multipartite} and
	\eqref{eq:relative_final} into Eq.~\eqref{eq:ergotropy_relation}, the
	scaled ergotropy is finally expressed as
	\begin{equation}
		\begin{aligned}
			\beta\epsilon(\rho(t))
			&=
			\sum_{X=A,B,C,\ldots}S(\rho_X)-S(\rho)
			-
			\sum_i\lambda_i\ln\lambda_i
			+
			\sum_{i,k}
			\lambda_i
			|\langle e_i|\phi_k\rangle|^2
			\ln\beta_k .
		\end{aligned}
		\label{eq:ergotropy_final}
	\end{equation}
	Here, $\lambda_i$ are the eigenvalues associated with the passive state
	$P_{\rho}$, while $\beta_k$ are the eigenvalues of the product state
	$\bigotimes_X\rho_X$ and, in general, $\beta_k\neq\lambda_i$.
\section{Numerical \textbf{results and discussions}}
In this section we present our results. For numerical studies, we evaluate the system dynamics using a time step of $0.01$ throughout the evolution. At each time step, we calculate the multipartite quantum mutual information and quantum relative entropy from time dependent density matrix of our system. The corresponding passive state is also constructed numerically at each time step. These quantities are then used to determine the dynamical behavior of the quantum ergotropy and to study how the available extractable work changes with system size, thermal photon number, and the choice of the initial atomic state.
	
\begin{figure}[H]
		\centering
		\includegraphics[width=5.5in]{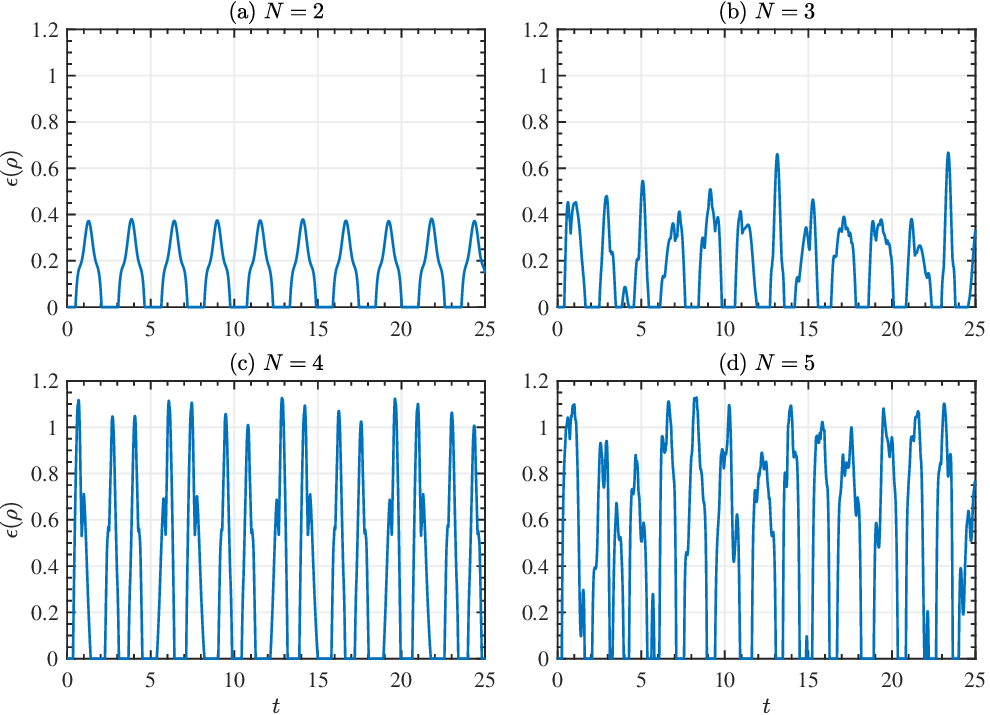}
		\caption{(color online) Dynamics of the quantum ergotropy $\epsilon(\rho)$ for different numbers of two-level atoms: (a) $N=2$, (b) $N=3$, (c) $N=4$, and (d) $N=5$. The results show the change in both the maximum extractable work and the intervals of vanishing ergotropy as the system size increases. We take $\bar{n}=1$, $p=0.5$ and $\alpha=3\pi/4$.}
		\label{fig1}
	\end{figure}

\begin{figure}[H]
		\centering
		\includegraphics[width=5.5in]{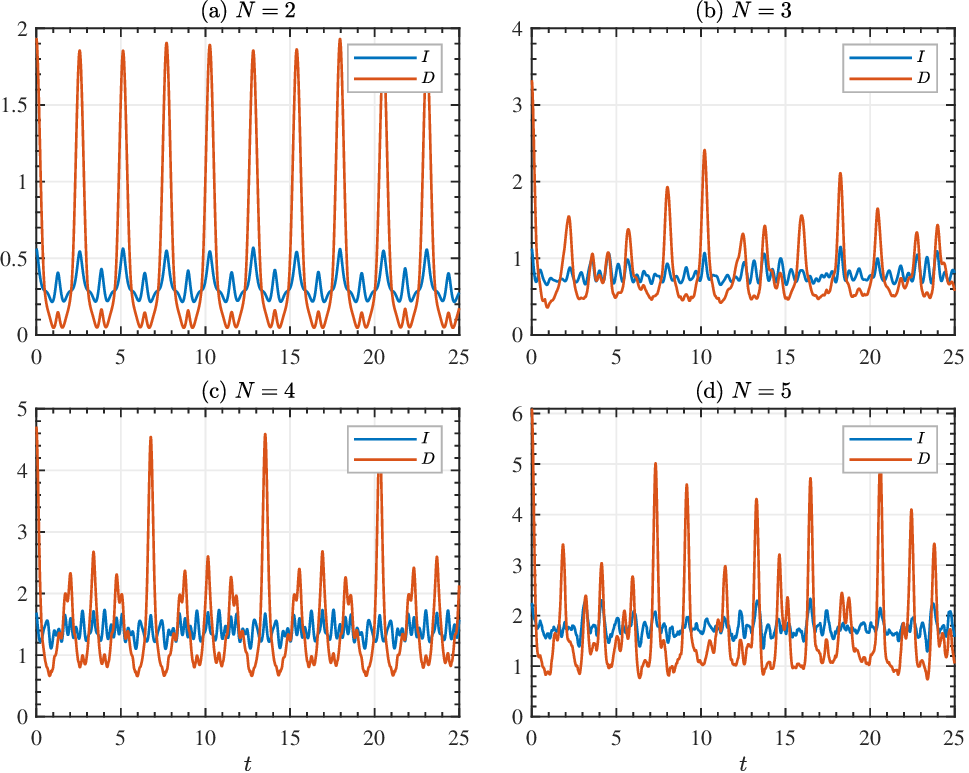}
		\caption{(color online) Time evolution of the multipartite quantum mutual information $I$ and quantum relative entropy $D$ for different numbers of two-level atoms: (a) $N=2$, (b) $N=3$, (c) $N=4$, and (d) $N=5$. The relative entropy represents the distinguishability between the passive state and the tensor product of the local reduced states. All the other parameters are same as for Fig. (\ref{fig1}).}
		\label{fig2}
	\end{figure}

\begin{figure}[H]
		\centering
		\includegraphics[width=5.5in]{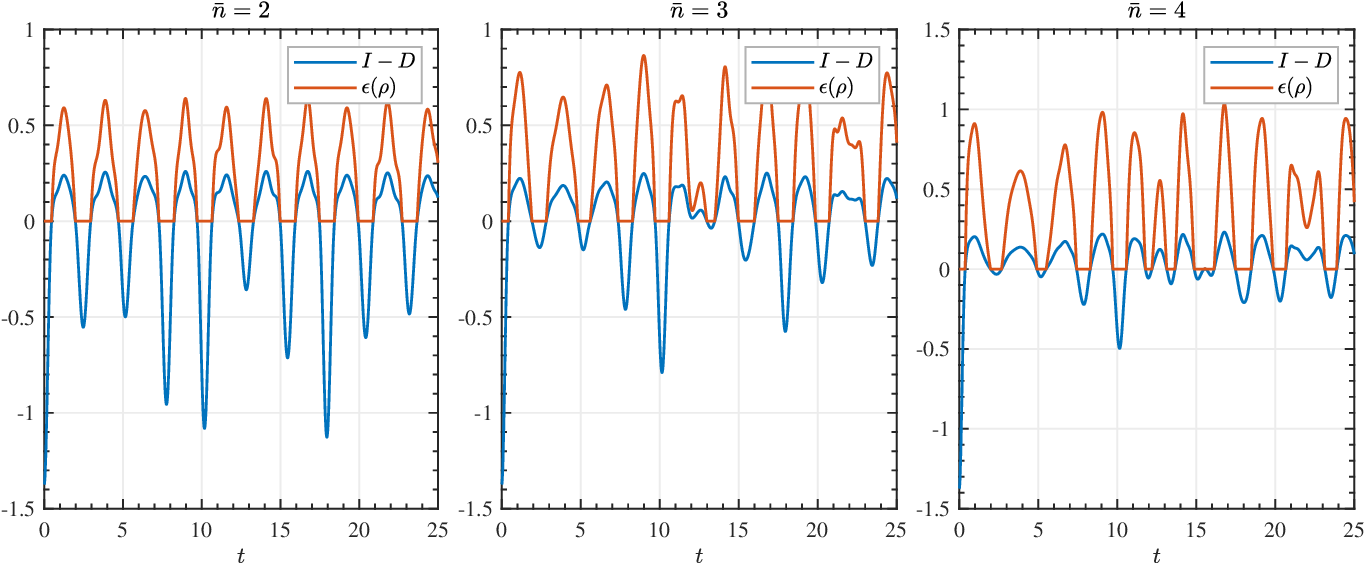}
		\caption{(color online) Dynamics of $I-D$ and quantum ergotropy $\epsilon(\rho)$ for different average thermal photon numbers: (a) $\bar{n}=2$, (b) $\bar{n}=3$, and (c) $\bar{n}=4$. The system under consideration is $N=2$ and the other parameter are fixed as for Fig. (\ref{fig1}). The figure illustrates how thermal photons modify the balance between the mutual-information and relative-entropy contributions and, consequently, the available extractable work.}
		\label{fig3}
	\end{figure}

\begin{figure}[H]
		\centering
		\includegraphics[width=5.5in]{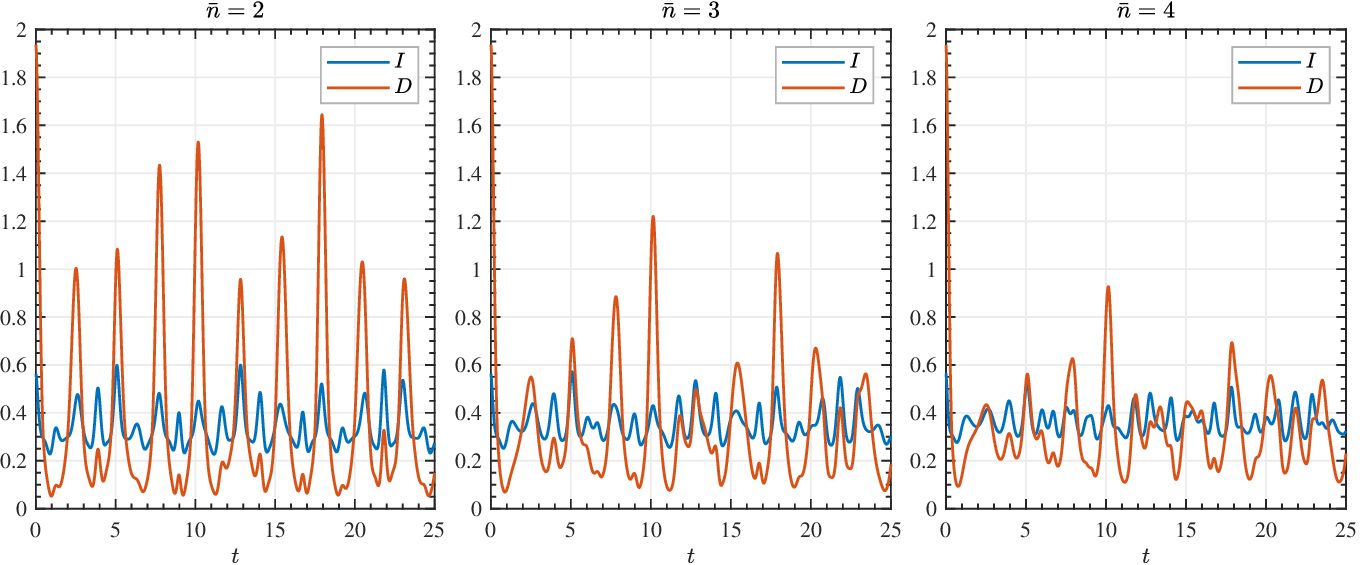}
		\caption{(color online) Time evolution of the quantum mutual information $I$ and quantum relative entropy $D$ for different average thermal photon numbers: (a) $\bar{n}=2$, (b) $\bar{n}=3$, and (c) $\bar{n}=4$. The system we take is $N=2$, and all the other parameter are same as Fig. (\ref{fig1}). Increasing the thermal photon number modifies the relative-entropy contribution while the mutual information remains comparatively robust.}
		\label{fig4}
	\end{figure}

\begin{figure}[H]
		\centering
		\includegraphics[width=5.5in]{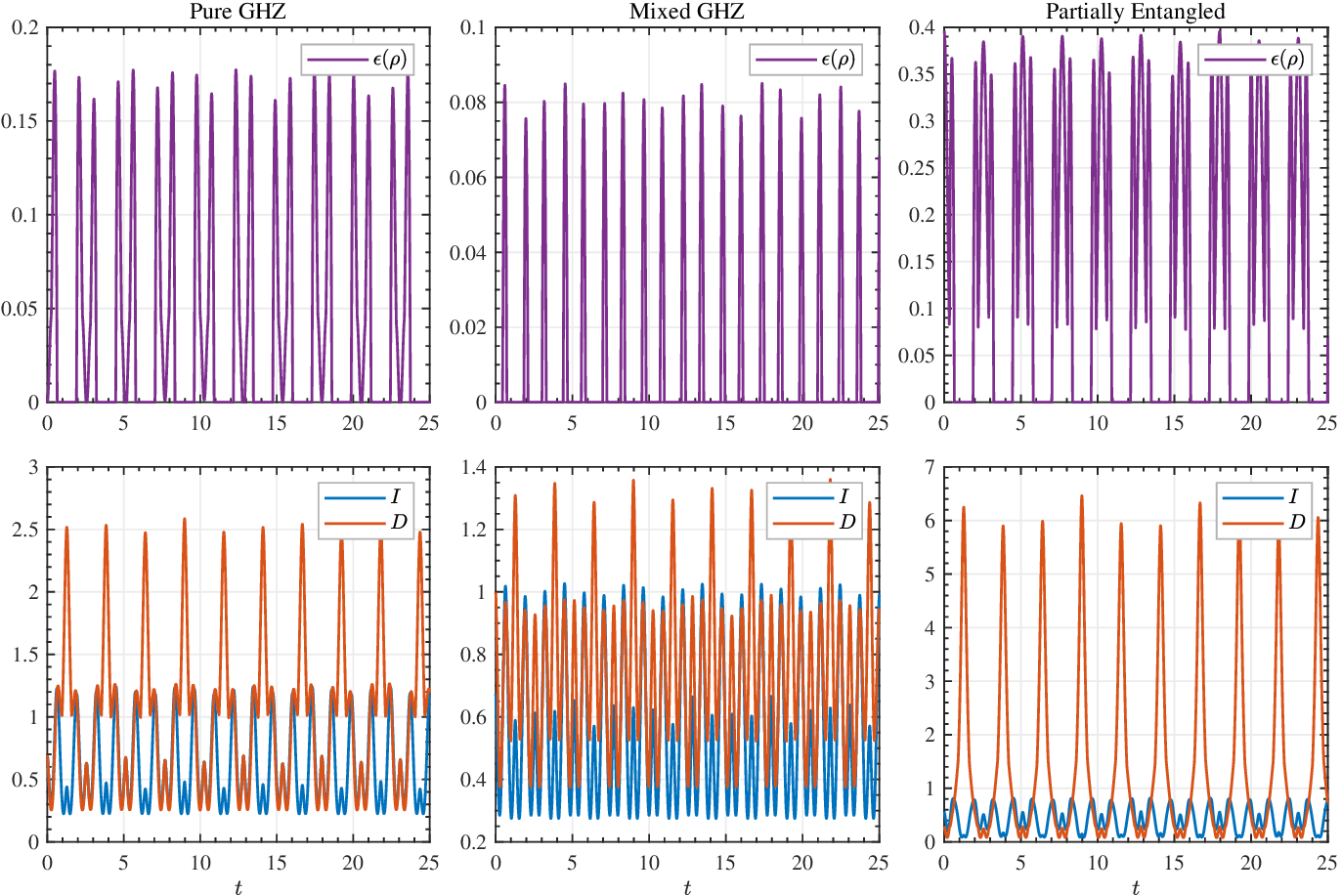}
		\caption{(color online) Dynamics of quantum ergotropy $\epsilon(\rho)$ (upper row) and the corresponding quantum mutual information $I$ and relative entropy $D$ (lower row) for different initial atomic states: pure GHZ state (left column) with $p=0$ and $\alpha=\pi/4$, mixed GHZ state (middle column) with $p=0.2$ and $\alpha=\pi/4$, and partially entangled state (right column) with $p=0$ and $\alpha=\pi/6$. The results demonstrate the strong dependence of extractable work and correlation dynamics on the choice of the initial quantum state. We fix $N=2$ and $\bar{n}=1$.}
		\label{fig5}
	\end{figure}

The dynamics shown in Fig. (\ref{fig1}) reveal a clear change in the behavior of quantum ergotropy as the number of two-level atoms is increased. For $N=2$, the ergotropy exhibits a regular oscillatory behavior. It shows a finite work-extraction intervals followed by well-defined regions where the ergotropy vanishes. This alternating behavior indicates that the correlations generated during the evolution are periodically converted into useful work and then become ineffective for work extraction. When the system is extended to $N=3$, the dynamics become noticeably more irregular and the maximum ergotropy is enhanced. It reflects a richer structure of correlations available in the multipartite system. For $N=4$, the enhancement becomes more pronounced and the intervals of zero ergotropy become narrower. For this case the dynamics show that nonzero extractable work survives over a larger part of the evolution. This behavior is naturally understood from the competition between the multipartite mutual information and the relative-entropy contribution in Eq. (\ref{eq:ergotropy_relation}). The mutual information represents the correlations shared among the atomic subsystems and acts as a resource for generating finite ergotropy. On the other hand, the relative entropy quantifies how distinguishable the passive state is from the tensor product of the local reduced states. A larger distinguishability contributes negatively to the available ergotropy and therefore limits the amount of correlations that can actually be converted into useful work. Consequently, a large amount of multipartite correlation does not necessarily lead to large ergotropy. At times when the ergotropy decreases or becomes zero, the relative-entropy contribution becomes sufficiently large to compensate for the work-generating effect of the mutual information. The vanishing ergotropy can therefore be viewed as a balance between the correlations present in the system and the mismatch between the passive state and the corresponding product of local states. For $N=5$, the maximum ergotropy does not show a further significant enhancement compared with $N=4$. Instead, the main effect of the additional atom is the further suppression of the zero-ergotropy intervals. Thus, increasing the system size does not simply increase the peak extractable work; rather, it changes how the available correlations are distributed during the evolution and how strongly they are counteracted by the relative-entropy contribution. The overall behavior therefore demonstrates a non-linear dependence of work extraction on the number of atoms, with larger multipartite systems providing a more persistent, rather than necessarily larger, ergotropy.
Fig. (\ref{fig2}) shows the dynamics of the multipartite quantum mutual information $I$ and the quantum relative entropy $D$ for different numbers of two-level atoms. For $N=2$, both quantities exhibit a regular and strongly periodic behavior. The mutual information remains comparatively small, oscillating mainly between about $0.2$ and $0.6$, whereas the relative entropy shows much larger oscillations, with sharp peaks approaching $1.9$. The dominance of $D$ over $I$ during several time intervals explains the appearance of zero-ergotropy regions in Fig. (\ref{fig1}), since the relative-entropy contribution can compensate for the work-generating role of the correlations. For $N=3$, the dynamics become more irregular. The mutual information is enhanced and typically remains around $0.6$--$1.1$, while the relative entropy fluctuates over a wider range and reaches values of about $2.4$. Interestingly, $D$ also develops pronounced oscillations and sharp peaks, showing that the difference between passive state and the product of the local states becomes increasingly important. Nevertheless, there are time intervals where $I$ becomes comparable to or larger than $D$, allowing a finite ergotropy to emerge. For $N=4$, both $I$ and $D$ are further enhanced, with $I$ reaching approximately $1.8$ and $D$ exhibiting peaks as high as $4.5$. Although the multipartite correlations become stronger, they are not converted into work in a straightforward manner, because the increasing relative-entropy contribution can strongly suppress the resulting ergotropy. The dynamics nevertheless provide more frequent intervals in which the balance between $I$ and $D$ favors finite work extraction. For $N=5$, the mutual information remains at a relatively high level, reaching values around $2.2$, while the relative entropy shows even stronger fluctuations with peaks approaching $5$. This demonstrates that enlarging the atomic system enhances both the available correlations and the distinguishability contribution. The figure therefore provides a direct explanation for the nontrivial ergotropy dynamics: finite ergotropy appears when the mutual-information contribution dominates the relative-entropy penalty, whereas a large $D$ suppresses or completely eliminates the extractable work. Hence, increasing the number of atoms does not simply increase the mutual information and, consequently, the ergotropy; rather, it strengthens the competition between the two contributions and changes the temporal windows in which useful work can be extracted.
Fig. (\ref{fig3}) illustrates how the dynamics of the net correlation contribution, $I-D$, and the quantum ergotropy $\epsilon(\rho)$ are modified by increasing the average thermal photon number $\bar{n}$. For $\bar{n}=2$, both quantities exhibit a clear and regular oscillatory structure. The ergotropy remains nonzero over well-defined time intervals, while $I-D$ alternates between positive and negative values. In particular, the negative regions of $I-D$ correspond to times when the relative-entropy contribution dominates the mutual information, resulting in vanishing ergotropy. When the thermal photon number is increased to $\bar{n}=3$, the regularity of the dynamics is partially disturbed. The oscillations of both $I-D$ and $\epsilon(\rho)$ become more irregular, and the maximum ergotropy increases from its value at $\bar{n}=2$. At the same time, the negative excursions of $I-D$ are reduced in magnitude over several time intervals, indicating that the relative-entropy penalty becomes less dominant at those times. Consequently, the system maintains finite ergotropy over a broader range of the evolution. For $\bar{n}=4$, the amplitude of $I-D$ is generally reduced compared with $\bar{n}=2$ and $\bar{n}=3$, while the ergotropy remains finite but develops a more irregular temporal structure. The reduction of $I-D$ reflects the fact that increasing thermal photons does not simply enhance the correlation resource available for work extraction. Instead, thermal fluctuations modify both the mutual information and the distinguishability contribution, leading to a redistribution of the time intervals in which useful work can be extracted. Overall, the figure shows that increasing $\bar{n}$ does not produce a monotonic enhancement of ergotropy. Thermal photons substantially alter the balance between $I$ and $D$, and hence the net quantity $I-D$ that determines the available work. The resulting dynamics demonstrate that thermal excitation can enhance ergotropy at certain times while suppressing it at others, leading to a more irregular but temporally redistributed work-extraction process as $\bar{n}$ increases.
Fig. (\ref{fig4}) shows the influence of the average thermal photon number $\bar{n}$ on the dynamics of the quantum mutual information $I$ and the relative entropy $D$. For $\bar{n}=2$, the mutual information remains relatively stable, with oscillations mainly in the range of about $0.25$--$0.60$. In contrast, $D$ exhibits much stronger variations, with several sharp peaks reaching values above $1.5$. As the average thermal photon number increases to $\bar{n}=3$, the mutual information changes only moderately and remains within approximately the same range. However, a noticeable suppression of the relative entropy is observed, with its largest dynamical peaks decreasing to about $1.2$. This tendency becomes more evident for $\bar{n}=4$, where the strong peaks of $D$ are further reduced and its dynamics become closer in magnitude to those of $I$. The largest value of $D$ during the evolution is then below approximately $1$, apart from the initial value. These results show that increasing the thermal photon number affects $D$ more strongly than $I$. 
Consequently, the negative contribution of the relative entropy to work extraction becomes weaker. At the same time, the mutual information remains comparatively robust, indicating that a considerable amount of correlation between the atomic subsystems survives in the presence of additional thermal photons. The decreasing separation between $I$ and $D$ with increasing $\bar{n}$ therefore explains the corresponding modification of the ergotropy dynamics: the relative-entropy contribution competes less strongly with the correlation resource, thereby reducing the intervals in which the available correlations are completely cancelled for work extraction.
Fig. (\ref{fig5}) compares the dynamics of quantum ergotropy $\epsilon(\rho)$, quantum mutual information $I$, and quantum relative entropy $D$ for three different initial atomic states: the pure GHZ state, the mixed GHZ state, and a partially entangled state. A clear dependence of the work-extraction dynamics on the nature of the initial state is observed. For the pure GHZ state, the ergotropy shows an almost regular oscillatory behavior, with repeated intervals of finite and zero extractable work. Its maximum value is about $0.18$. The corresponding mutual information and relative entropy also display regular oscillations. The relative entropy shows several pronounced peaks reaching values of approximately $2.5$, while the mutual information remains comparatively smaller. The intervals in which $D$ becomes dominant are associated with the suppression of ergotropy, whereas finite ergotropy appears when the correlation contribution is sufficiently large to overcome the relative-entropy contribution. For the mixed GHZ state, the maximum ergotropy is considerably reduced and remains below approximately $0.09$. The dynamics are also more rapid, with narrow intervals of finite ergotropy separated by frequent zero regions. Compared with the pure GHZ case, both $I$ and $D$ remain within a smaller range for most of the evolution, although $D$ still develops isolated peaks of about $1.3$--$1.4$. The relatively small ergotropy indicates that the correlations present in the mixed state are less effective for work extraction. In this case, the competition between $I$ and $D$ more frequently results in a small net contribution available for conversion into useful work. A significantly different behavior is obtained for the partially entangled state. The ergotropy is strongly enhanced and reaches values close to $0.4$, which is more than twice the maximum obtained for the pure GHZ state. Finite ergotropy also appears over broader portions of the evolution, although narrow zero-ergotropy intervals are still present. The mutual information remains finite and oscillatory, while the relative entropy exhibits very large and sharp peaks, reaching values above $6$ at selected times. Away from these sharp peaks of relative entropy, the balance becomes more favorable to the correlation contribution and comparatively large ergotropy is recovered. The comparison therefore shows that the initial character of the quantum state plays an important role in determining both the amount and the persistence of extractable work. Among the three cases, the partially entangled state is the most favorable for obtaining larger and more persistent nonzero ergotropy, whereas the pure GHZ state provides a smaller but more regular work-extraction pattern. The mixed GHZ state is the least favorable for maximum work extraction because its ergotropy remains substantially lower and vanishes more frequently during the evolution.
\section{Conclusion}

In this work, we have investigated the dynamics of quantum ergotropy, multipartite quantum mutual information, and quantum relative entropy in a system of interacting two-level atoms. The main aim was to understand how the extractable work changes with the number of atoms and how it is affected by multipartite correlations, thermal photons, and the choice of the initial atomic state. In this picture, the mutual information acts as the correlation resource, while the relative entropy measures the distinguishability between the passive state and the tensor product of the local reduced states. Their competition determines whether the system can provide finite extractable work during the evolution.

Our numerical results show that the ergotropy changes strongly with the system size, but not in a simple linear way. For $N=2$, the dynamics are regular, with repeated finite-ergotropy and zero-ergotropy intervals. For $N=3$ and $N=4$, the maximum ergotropy is enhanced and the dynamics become more irregular, while the zero-ergotropy intervals are reduced. For $N=5$, the maximum ergotropy does not increase much further, but the zero regions become more suppressed. The separate behavior of $I$ and $D$ helps to explain these features: when $D$ becomes large compared with the correlation contribution, the ergotropy decreases and can vanish. Increasing the average thermal photon number mainly suppresses the strong peaks of $D$, while $I$ remains comparatively stable. The initial state also plays an important role. The pure GHZ state gives regular work-extraction dynamics, the mixed GHZ state produces the smallest ergotropy, while the partially entangled state gives the largest and more persistent nonzero ergotropy.

These results are also relevant to the study of quantum batteries, where ergotropy, charging power, and stability of stored energy are important quantities. The present model can be extended to investigate charging and discharging processes, collective charging advantages, and the relation between multipartite correlations and quantum power. It would also be useful to study whether the suppression of zero-ergotropy intervals for larger systems leads to improved charging performance. A further extension should include cavity loss, atomic decay, and dephasing, in order to test how robust these work-extraction advantages remain under realistic open-system conditions.

\end{document}